arXiv:quant-ph/9808027v1 17 Aug 1998

# Parallel Quantum Computation and Quantum Codes

Cristopher Moore[1] and Martin Nilsson[2]

[1] Santa Fe Institute, 1399 Hyde Park Road, Santa Fe, New Mexico 87501
moore@santafe.edu

[2] Chalmers Tekniska Högskola and University of Göteborg, Göteborg, Sweden
martin@fy.chalmers.se

**Abstract.** We propose a definition of **QNC**, the quantum analog of the efficient parallel class **NC**. We exhibit several useful gadgets and prove that various classes of circuits can be parallelized to logarithmic depth, including circuits for encoding and decoding standard quantum error-correcting codes, or more generally any circuit consisting of controlled-not gates, controlled $\pi$-shifts, and Hadamard gates. Finally, while we note the Quantum Fourier Transform can be parallelized to linear depth, we conjecture that an even simpler 'staircase' circuit cannot be parallelized to less than linear depth, and might be used to prove that **QNC** $<$ **QP**.

## 1 Introduction

Much of computational complexity theory has focused on the question of what problems can be solved in polynomial time. Shor's quantum factoring algorithm [14] suggests that quantum computers might be more powerful than classical computers in this regard, i.e. that **QBP** might be a larger class than **P**, or rather **BP**, the class of problems solvable in polynomial time by a classical probabilistic Turing machine with bounded error.

A finer distinction can be made between **P** and the class **NC** of efficient parallel computation, namely the subset of **P** of problems which can be solved by a parallel computer with a polynomial number of processors in *polylogarithmic* time, $\mathcal{O}(\log^k n)$ time for some $k$, where $n$ is the number of bits of the input [12]. Equivalently, **NC** problems are those solvable by Boolean circuits with a polynomial number of gates and polylogarithmic depth.

This distinction seems especially relevant for quantum computers, where decoherence makes it difficult to do more than a limited number of computation steps reliably. Since decoherence due to storage errors is essentially a function of time, we can avoid it by doing as many of our quantum operations at once as possible; if we can parallelize our computation to logarithmic depth, we can solve exponentially larger problems. (Gate errors, on the other hand, will typically get worse, since parallel algorithms often involve more gates.)

In this paper, we propose a definition of **QNC** and prove a number of elementary results. Our main theorem is that circuits consisting of controlled-not gates, controlled $\pi$-shifts, and Hadamard gates can be parallelized to logarithmic depth. This includes circuits for encoding and decoding standard quantum error-correcting codes. We end with a conjecture that a simple 'staircase' circuit cannot be parallelized, and so might be used to prove that $\mathbf{QNC} < \mathbf{QP}$.

## 2 Definitions

We define quantum operators and quantum circuits as follows:

**Definition 1.** *A* quantum operator *on $n$ qubits is a unitary rank-$2n$ tensor $U$ where $U^{b_1 b_2 \ldots b_n}_{a_1 a_2 \ldots a_n}$ is the amplitude of the incoming and outgoing truth values being $a_1, a_2, \ldots a_n$ and $b_1, b_2, \ldots b_n$ respectively, with $a_i, b_i \in \{0, 1\}$ for all $i$. However, we will usually write $U$ as a $2^n \times 2^n$ unitary matrix $U_{ab}$ where $a$ and $b$'s binary representations are $a_1 a_2 \cdots a_n$ and $b_1 b_2 \cdots b_n$ respectively.*

*A* one-layer circuit *consists of the tensor product of one- and two-qubit gates, i.e. rank 2 and 4 tensors, or $2 \times 2$ and $4 \times 4$ unitary matrices. This is an operator that can be carried out by a set of simultaneous one-qubit and two-qubit gates, where each qubit interacts with at most one gate.*

*A* quantum circuit of depth *$k$ is a quantum operator written as the product of $k$ one-layer circuits.*

Here we are allowing arbitrary two-qubit gates. If we like, we can restrict this to *controlled-$U$ gates*, of the form $\begin{pmatrix} 1 & 0 & 0 & 0 \\ 0 & 1 & 0 & 0 \\ 0 & 0 & u_{11} & u_{12} \\ 0 & 0 & u_{21} & u_{22} \end{pmatrix}$, or more stringently to the *controlled-not* gate $\begin{pmatrix} 1 & 0 & 0 & 0 \\ 0 & 1 & 0 & 0 \\ 0 & 0 & 0 & 1 \\ 0 & 0 & 1 & 0 \end{pmatrix}$. For these, we will call the first and second qubits the *input* and *target* qubit respectively, even though they don't really leave the input qubit unchanged, since they entangle it with the target.

Since either of these can be combined with one-qubit gates to simulate arbitrary two-qubit gates [1], these restrictions would just multiply our definition of depth by a constant. The same is true if we wish to allow gates that couple $k > 2$ qubits as long as $k$ is fixed, since any $k$-qubit gate can be simulated by some constant number of two-qubit gates.

In order to design a shallow parallel circuit for a given quantum operator, we want to be able to use additional qubits or "ancillae" for intermediate steps in the computation, equivalent to additional processors in a parallel quantum computer. However, to avoid entanglement, we demand that our ancillae start and end in a pure state $|0\rangle$, so that the desired operator appears as the diagonal block of the operator performed by the circuit on the subspace where the ancillae are zero.

Then in analogy with $\mathbf{NC}$ we propose the following definition:

**Definition 2.** *Let $F$ be a family of quantum operators, i.e. $F(n)$ is a $2^n \times 2^n$ unitary matrix on $n$ qubits. We say that $F(n)$ is* embedded *in an operator $M$ with $m$ ancillae if $M$ is a $2^{m+n} \times 2^{m+n}$ matrix which preserves the subspace* *where the ancillae are set to* $|0\rangle$*, and if* $M$ *is identical to* $F(n) \otimes \mathbf{1^{2^m}}$ *when restricted to this subspace.*

*Then* $\mathbf{QNC} = \cup_k \mathbf{QNC}^k$ *where* $\mathbf{QNC}^k$ *is the class of operators parallelizable to* $\mathcal{O}(\log^k n)$ *depth with a polynomial number of ancillae. That is,* $F$ *is in* $\mathbf{QNC}^k$ *if, for some constants* $c_1$*,* $c_2$ *and* $j$*,* $F(n)$ *can be embedded in a circuit of depth at most* $c_1 \log^k n$*, with at most* $c_2 n^j$ *ancillae.*

To extend this definition from quantum operators to decision problems in the classical sense, we would have to choose a measurement protocol, and to what extent we want errors to be bounded. We will not explore those issues here.

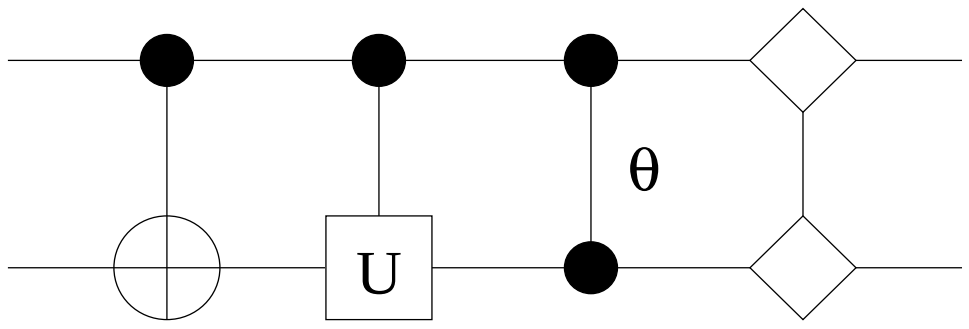

**Fig. 1.** Our notation for controlled-not, controlled-$U$, symmetric phase shift, and arbitrary diagonal gates.

We will use the notation in figure 1 for our various gates: the controlled-not and controlled-$U$, the symmetric phase shift $\begin{pmatrix} 1 & 0 & 0 & 0 \\ 0 & 1 & 0 & 0 \\ 0 & 0 & 1 & 0 \\ 0 & 0 & 0 & e^{i\theta} \end{pmatrix}$, and arbitrary diagonal gates $\begin{pmatrix} e^{i\omega_{00}} & 0 & 0 & 0 \\ 0 & e^{i\omega_{01}} & 0 & 0 \\ 0 & 0 & e^{i\omega_{10}} & 0 \\ 0 & 0 & 0 & e^{i\omega_{11}} \end{pmatrix}$.

A preliminary version of this work, lacking proposition 9 and all of section 7 on quantum codes, appeared as [11].

## 3 Permutations

In classical circuits, one can move wires around as much as one likes. In a quantum computer, it may be more difficult to move a qubit from place to place. However, we can easily do arbitrary permutations in constant depth:

**Proposition 1.** *Any permutation of* $n$ *qubits can be performed in 4 layers of controlled-not gates with* $n$ *ancillae, or in 6 layers with no ancillae.*

*Proof.* The first part is obvious; simply copy the qubits into the ancillae, cancel the originals, recopy them from the ancillae in the desired order, and cancel the ancillae. This is shown in figure 2.

Without ancillae, we can use the fact that any permutation can be written as the composition of two sets of disjoint transpositions [17]. To see this, first decompose it into a product of disjoint cycles, and then note that a cycle is the composition of two reflections, as shown in figure 3. Two qubits can be switched with 3 layers of controlled-not gates as shown in figure 4, so any permutation can be done in 6 layers. □

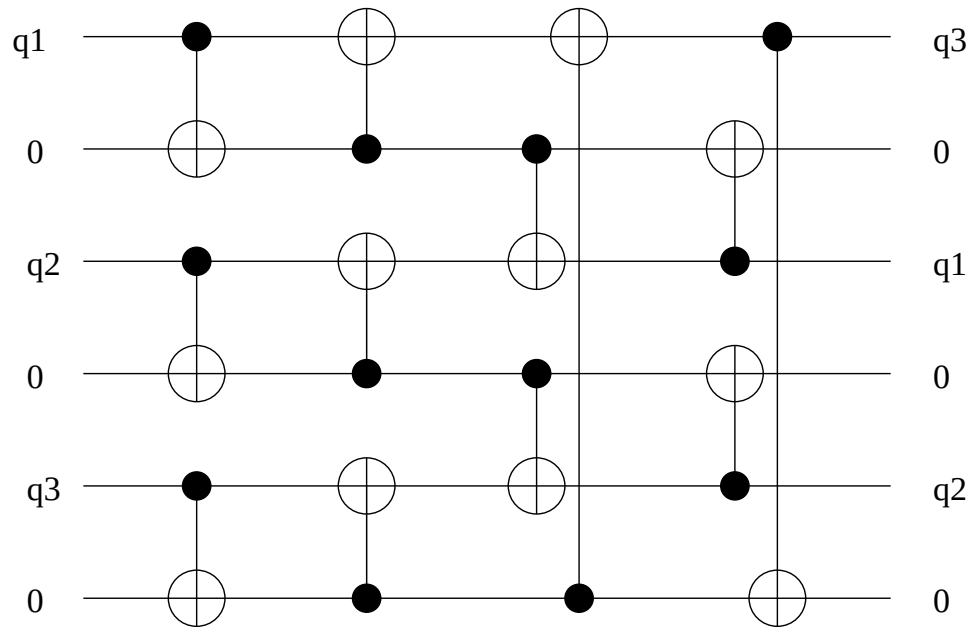

**Fig. 2.** Permuting $n$ qubits in 4 layers using $n$ ancillae.

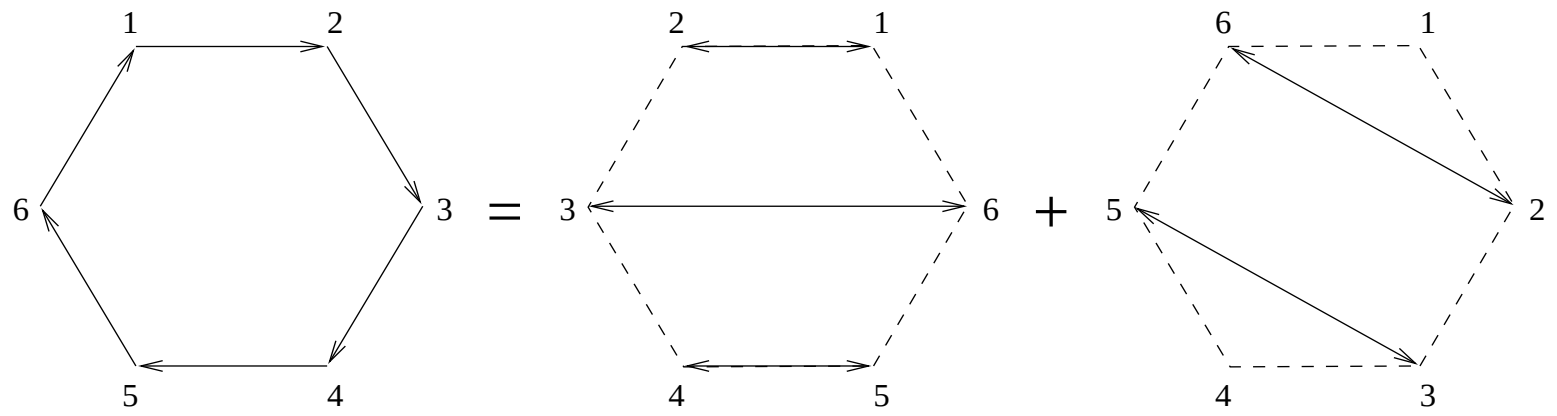

**Fig. 3.** Any cycle, and therefore any permutation, is the composition of two sets of disjoint transpositions.

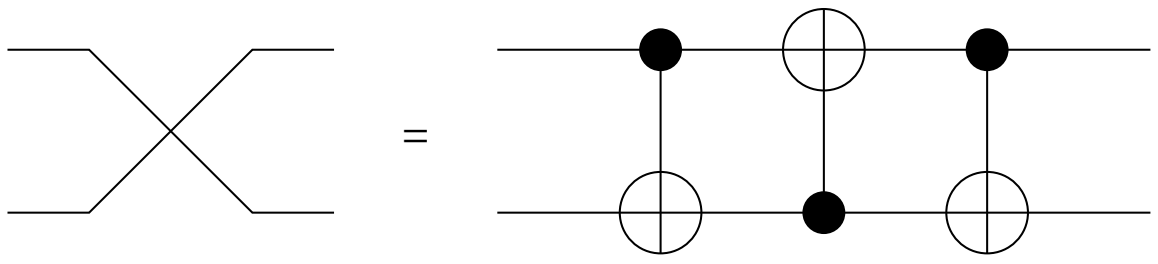

**Fig. 4.** Switching two qubits with three controlled-nots.

## 4 Fan-out

To make a shallow parallel circuit, it is often important to *fan out* one of the inputs into multiple copies. The controlled-not gate can be used to copy a qubit onto an ancilla in the pure state $|0\rangle$ by making a non-destructive measurement:

$$(\alpha|0\rangle + \beta|1\rangle) \otimes |0\rangle \;\to\; \alpha|00\rangle + \beta|11\rangle$$

Note that the final state is not a tensor product of two independent qubits, since the two qubits are completely entangled. Making an unentangled copy requires non-unitary, and in fact non-linear, processes since

$$(\alpha|0\rangle + \beta|1\rangle) \otimes (\alpha|0\rangle + \beta|1\rangle) = \alpha^2|00\rangle + \alpha\beta(|01\rangle + |10\rangle) + \beta^2|11\rangle$$

has coefficients quadratic in $\alpha$ and $\beta$. This is the classic 'no cloning' theorem.

This means that disentangling or *uncopying* the ancillae by the end of the computation, and returning them to their initial state $|0\rangle$, is a non-trivial and important part of a quantum circuit. There are, however, some special cases where this can be done easily.

Suppose we have a series of $n$ controlled-$U$ gates all with the same input qubit. Rather than applying them in series, we can *fan out* the input into $n$ copies by splitting it $\log_2 n$ times, apply them to the target qubits, and uncopy them afterward, thus reducing the circuit's depth to $\mathcal{O}(\log n)$ depth.

**Proposition 2.** *A series of $n$ controlled gates coupling the same input to $n$ target qubits can be parallelized to $\mathcal{O}(\log n)$ depth with $\mathcal{O}(n)$ ancillae.*

*Proof.* The circuit in figure 5 copies the input onto $n-1$ ancillae, applies all the controlled gates simultaneously, and uncopies the ancillae back to their original state. Its total depth is $2\log_2 n + 1$. □

This kind of symmetric circuit, in which we uncopy the ancillae to return them to their original state, is similar to circuits designed by the Reversible Computation Group at MIT [6] for reversible classical computers.

## 5 Diagonal and mutually commuting gates

Fan-in seems more difficult in general. Classically, we can calculate the composition of $n$ operators in $\mathcal{O}(\log n)$ time by composing them in pairs; but it is unclear when we can do this with unitary operators. One special case where it is possible is if all the gates are diagonal:

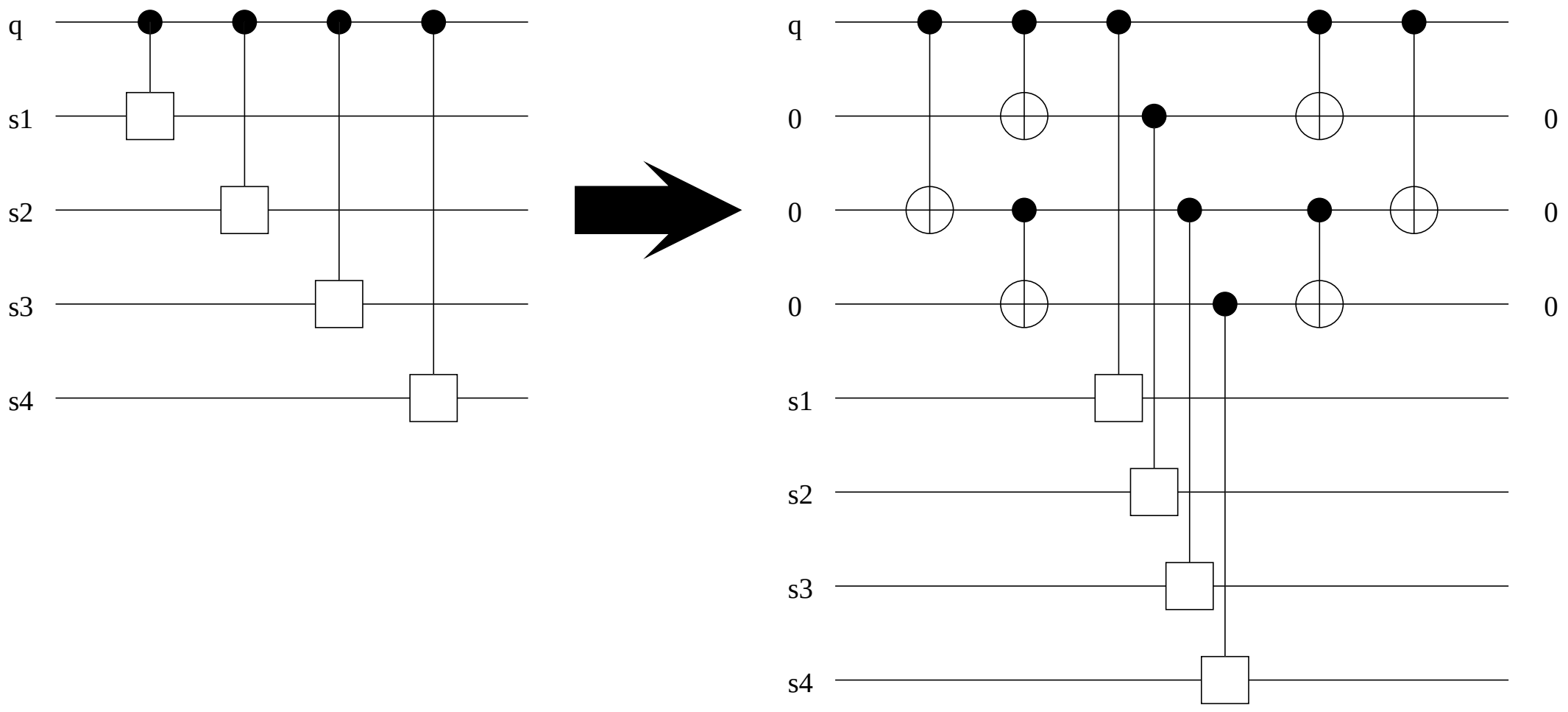

**Fig. 5.** Parallelizing $n$ controlled gates on a single input qubit $q$ to $\mathcal{O}(\log n)$ depth.

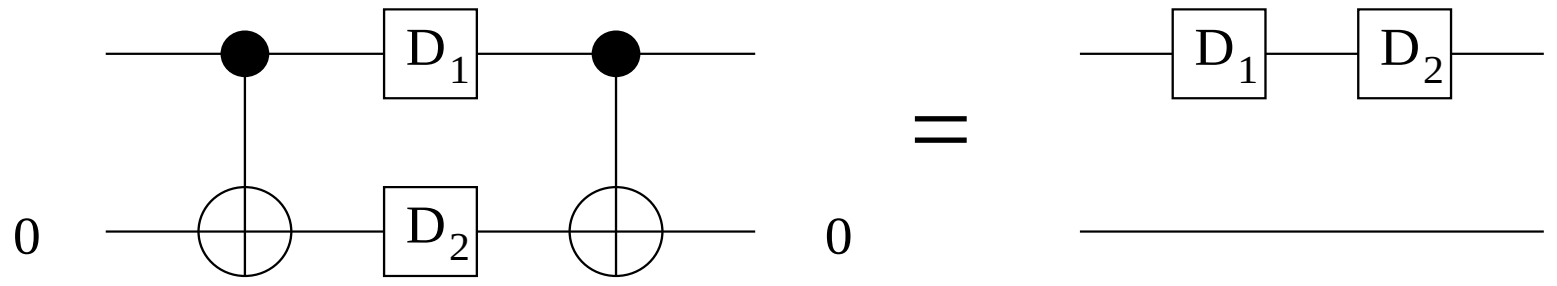

**Fig. 6.** Using entanglement to parallelize diagonal operators.

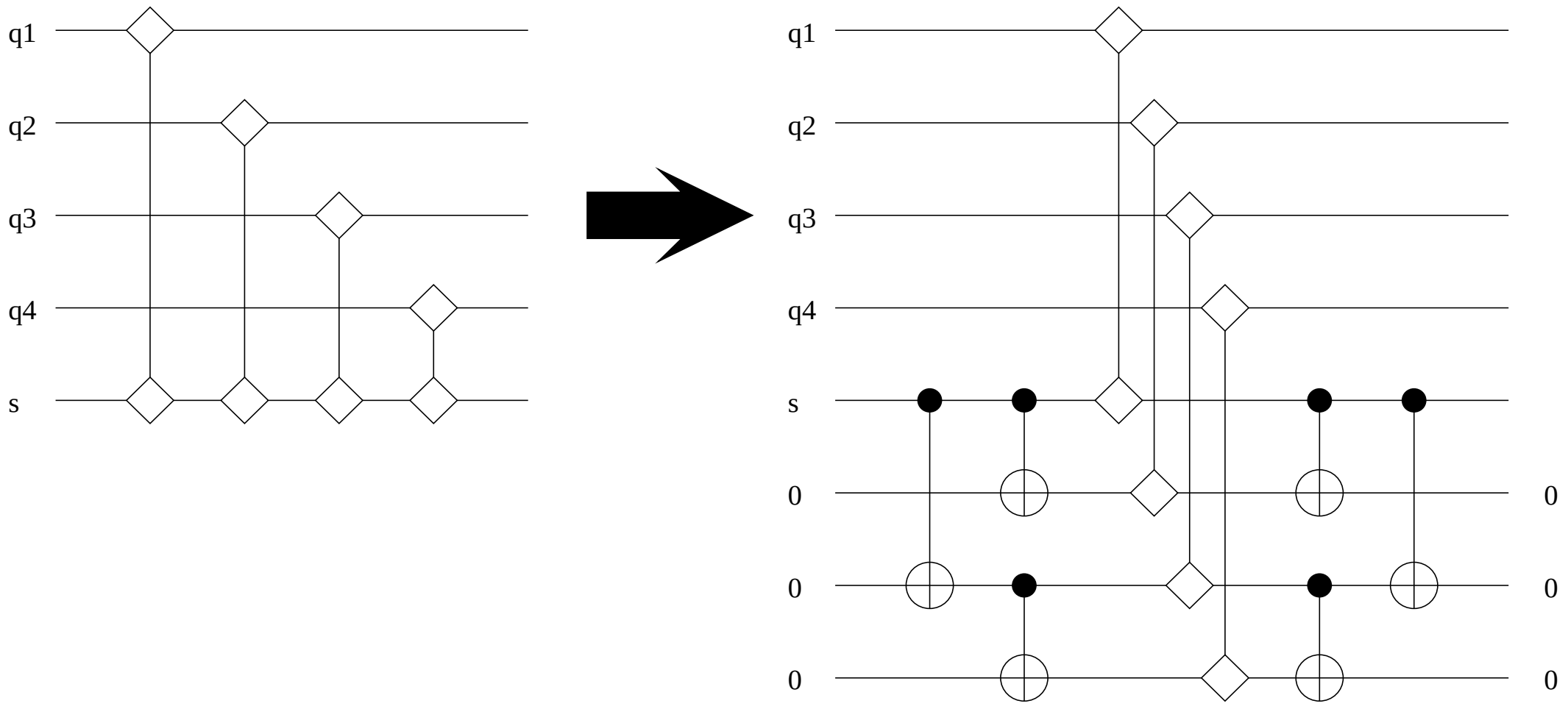

**Fig. 7.** Parallelizing $n$ diagonal gates on a single qubit as in proposition 2.

**Proposition 3.** *A series of $n$ diagonal gates on the same qubits can be parallelized to $\mathcal{O}(\log n)$ depth with $\mathcal{O}(n)$ ancillae.*

*Proof.* Here the entanglement between two copies of a qubit becomes an asset. Since diagonal matrices don't mix Boolean states with each other, we can act on one or more qubits and an entangled copy of them with two diagonal matrices $D_1$ and $D_2$ as in figure 6. When we uncopy the ancilla(e), we have the same effect as if we had applied both matrices to the original qubit(s). Then the same kind of circuit as in proposition 2 works, as shown in figure 7. □

Since matrices commute if and only if they can be simultaneously diagonalized, we can generalize this to the case where a set of controlled-$U$ gates applied to the same target qubit(s) have mutually commuting $U$'s:

**Proposition 4.** *A series of of $n$ controlled-$U$ gates acting on the same target qubit(s) where the $U$'s mutually commute can be parallelized to $\mathcal{O}(\log n)$ depth with $\mathcal{O}(n)$ ancillae.*

*Proof.* Since the $U$'s all commute, they can all be diagonalized by the same unitary operator $T$. Apply $T^\dagger$ to the target qubit(s), parallelize the circuit using proposition 3, and put the target qubit(s) back in the original basis by applying $T$. This is all done with a circuit of depth $2\log_2 n + 3$. □

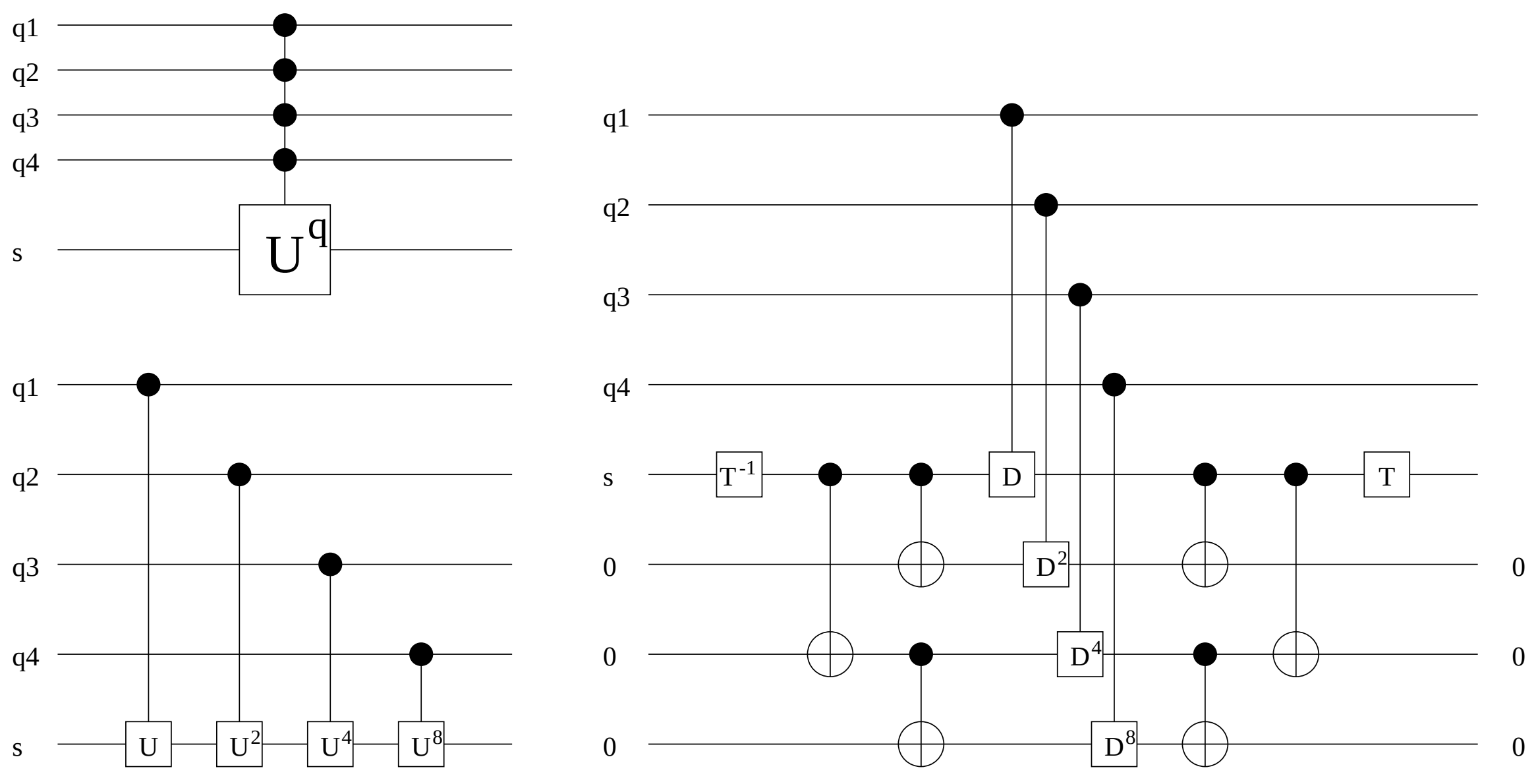

**Fig. 8.** Applying an operator $U$ $q$ times, where $q$ is given in binary by the input qubits.

As an example, in figure 8 we show a circuit that applies the $q$th power of an operator $U$ to a target qubit, where $0 \leq q < 2^k$ is given by $k$ input qubits as a binary integer. We can do this because $U, U^2, U^4, \ldots$ can be simultaneously diagonalized, since $U^q = T^\dagger D^q T$.

We can extend this to circuits in general whose gates are mutually commuting, which includes diagonal gates:

**Proposition 5.** *A circuit of any size consisting of diagonal or mutually commuting gates, each of which couples at most $k$ qubits, can be parallelized to depth $\mathcal{O}(n^{k-1})$ with no ancillae, and to depth $\mathcal{O}(\log n)$ with $\mathcal{O}(n^k)$ ancillae. Therefore, any family of such circuits is in* $\mathbf{QNC}^1$.

*Proof.* Since all the gates commute, we can sort them by which qubits they couple, and arrive at a compressed circuit with one gate for each $k$-tuple. This gives $\binom{n}{k} = \mathcal{O}(n^k)$ gates, but by performing groups of $n/k$ disjoint gates simultaneously we can do all of them in depth $\mathcal{O}(n^{k-1})$.

By making $\frac{k}{n}\binom{n}{k} = \mathcal{O}(n^{k-1})$ copies of each qubit, we can apply each gate to a disjoint set of copies as in propositions 3 and 4 to reduce this further to $\mathcal{O}(\log n)$ depth. □

This is hardly surprising; after all, diagonal gates commute with each other, which is almost like saying that they can be performed simultaneously.

## 6 Circuits of controlled-not gates

We can also fan in controlled-not gates. Figure 9 shows how to implement $n$ controlled-not gates on the same target qubit in depth $2\log_2 n + 1$. The ancillae carry the intermediate exclusive-ors of the inputs, and we combine them in pairs.

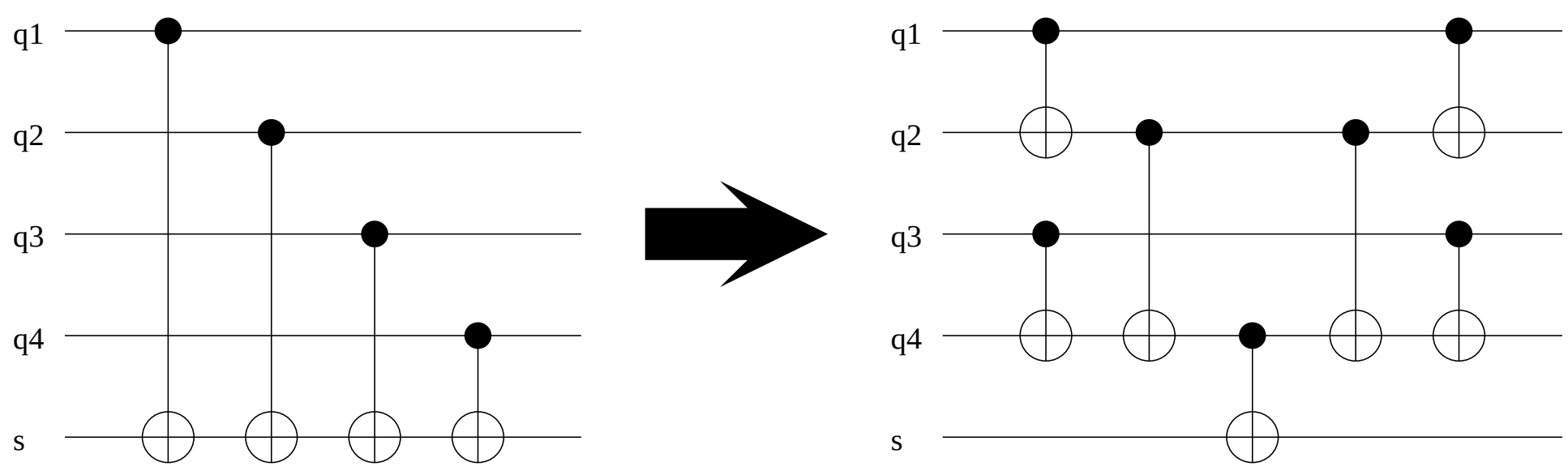

**Fig. 9.** Parallelizing $n$ controlled-not gates to $\mathcal{O}(\log n)$ depth by adding them in pairs.

We can use a generalization of this circuit to show that any circuit composed entirely of controlled-not gates can be parallelized to logarithmic depth:

**Proposition 6.** *A circuit of any size on $n$ qubits composed entirely of controlled-not gates can be parallelized to $\mathcal{O}(\log n)$ depth with $\mathcal{O}(n^2)$ ancillae. Therefore, any family of such circuits is in* $\mathbf{QNC}^1$.

*Proof.* First, note that in any circuit of controlled-not gates, if the $n$ input qubits have binary values and are given by an $n$-dimensional vector $q$, then the output can be written $Mq$ where $M$ is an $n \times n$ matrix over the integers mod 2. Each of the output qubits can be written as a sum of up to $n$ inputs, $(Mq)_i = \sum_k q_{j_k}$ where $j_k$ are those $j$ for which $M_{ij} = 1$.

We can break these sums down into binary trees. Let $W_n$ be the complete output sums, $W_{n/2}$ be their left and right halves consisting of up to $n/2$ inputs, and so on down to single inputs. There are less than $n^2$ such intermediate sums $W_k$ with $k > 1$. We assign an ancilla to each one, and build them up from the inputs in $\log_2 n$ stages, adding pairs from $W_k$ to make $W_{2k}$. The first stage takes $\mathcal{O}(\log n)$ time and an additional $\mathcal{O}(n^2)$ ancillae since we may need to make $\mathcal{O}(n)$ copies of each input, but each stage after that can be done in depth 2.

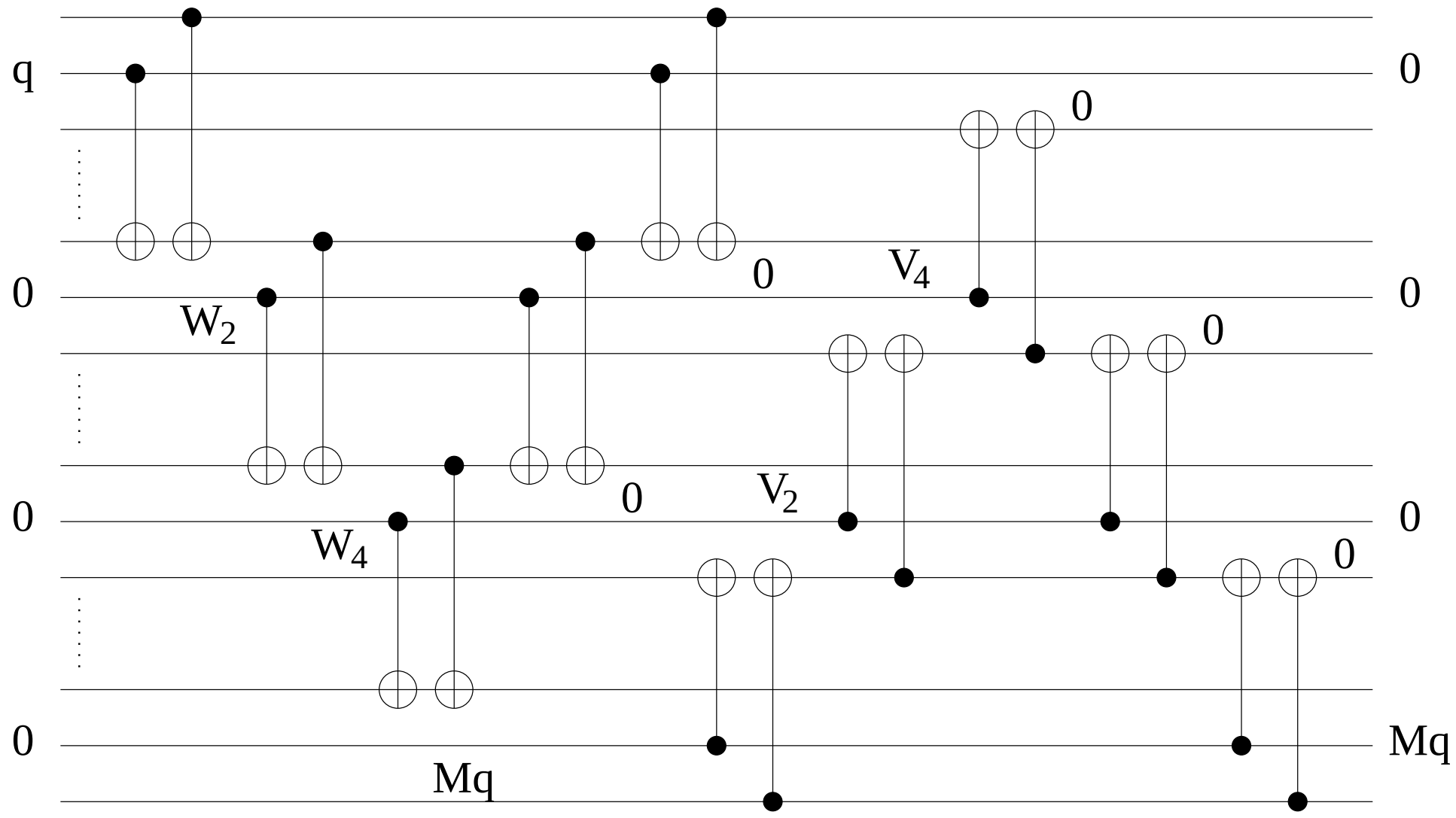

**Fig. 10.** Parallelizing an arbitrary circuit of controlled-not gates to logarithmic depth.

To cancel the ancillae, we use the same cascade in reverse order, adding pairs from $W_k$ to cancel $W_{2k}$. This leaves us with the input $q$, the output $Mq$, and the ancillae set to zero.

Now we use the fact that, since the circuit is unitary, $M$ is invertible. Thus we can recalculate the input $q = M^\dagger(Mq)$ and cancel it. We use the same ancillae in reverse order, building the inputs $q$ out of $Mq$ with a series of partial sums $V_2, V_4, \ldots$, cancel $q$, and cancel the ancillae in reverse as before. All this is illustrated in figure 10.

This leaves us with the output $Mq$ and all other qubits zero. With four more layers as in proposition 1, we can shift the output back to the input qubits, and we're done. □

This result is hardly surprising; after all, these circuits are reversible Boolean circuits, and any classical circuit composed of controlled-not gates is in $\mathbf{NC}^1$ (in fact, in the class $\mathbf{ACC}^0[2]$ of constant-depth circuits with sum mod 2 gates of unbounded fan-in). We just did a little extra work to disentangle the ancillae.

## 7 Controlled-not gates and phase shifts

We have shown that circuits composed of diagonal or controlled-not gates can be parallelized. It's reasonable to ask whether propositions 5 and 6 can be combined; that is, whether arbitrary circuits composed of controlled-not gates and diagonal operators can be parallelized to logarithmic depth. In this section, we will show that this is not the case.

**Proposition 7.** *Any diagonal unitary operator on $n$ qubits can be performed by a circuit consisting of an exponential number of controlled-not gates and one-qubit diagonal gates and no ancillae.*

*Proof.* Any diagonal unitary operator on $n$ qubits consists of $2^n$ phase shifts, $\begin{pmatrix} e^{i\omega_0} & & \\ & \ddots & \\ & & e^{i\omega_{2^n-1}} \end{pmatrix}$. If we write the phase angles as a $2^n$-dimensional vector $\omega$, then the effect of composing two diagonal operators is simply to add these vectors mod $2\pi$.

For each subset $s$ of the set of qubits, define a vector $\mu_s$ as $+1$ if the number of true qubits in $s$ is even, and $-1$ if it is odd. If $s$ is all the qubits, for instance, $\mu_{\{1\ldots n\}}$ is the aperiodic Morse sequence $(+1, -1, -1, +1, \ldots)$ when written out linearly, but it really just means giving the odd and even nodes of the Boolean $n$-cube opposite signs.

It is easy to see that the $\mu_s$ for all $s \subset \{1, \ldots, n\}$ are linearly independent, and form a basis of $\mathbb{R}^{2^n}$. Moreover, while diagonal gates coupling $k$ qubits can only perform phase shifts spanned by those $\mu_s$ with $|s| \leq k$, the circuit in figure 11 can perform a phase shift proportional to $\mu_s$ for any $s$ (incidentally, in depth $\mathcal{O}(\log |S|)$ with no ancillae). Therefore, a series of $2^n$ such circuits, one for each subset of $\{1, \ldots, n\}$, can express any diagonal unitary operator. □

This exponential bound is necessary in the worst case:

**Proposition 8.** *There are diagonal operators that cannot be parallelized to less than exponential depth with a polynomial number of ancillae.*

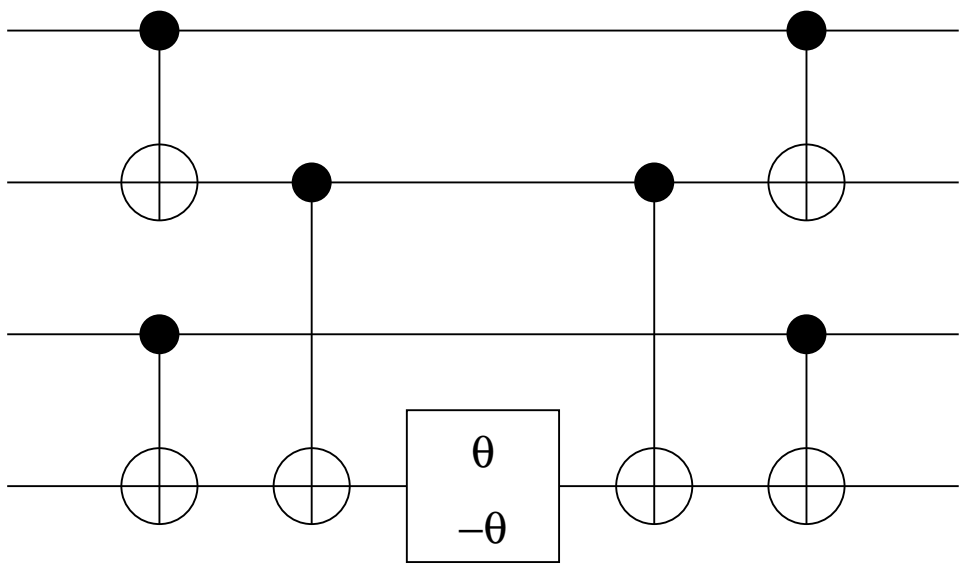

**Fig. 11.** A circuit for the phase shift $\theta\mu_s$, i.e. a phase shift of $+\theta$ if the number of true qubits is even and $-\theta$ if it is odd.

*Proof.* Consider setting up a many-to-one correspondence between circuits and operators. The set of diagonal unitary operators on $n$ qubits has $2^n$ continuous degrees of freedom, while the set of circuits of depth $d$ with $m$ ancillae has only $\mathcal{O}(d(m+n))$ continuous degrees of freedom (and some discrete ones for the circuit's topology). Thus if $m$ is polynomial, $d$ must be exponential. □

However, the next proposition shows that this won't help us distinguish **QP** from **QNC**. In fact, for controlled-nots and diagonal gates, **QP** and **QNC** are identical:

**Proposition 9.** *Any circuit consisting of controlled-not gates and $m$ diagonal operators coupling $k$ qubits each can be parallelized to $\mathcal{O}(\log n)$ depth with $\mathcal{O}(\max(kmn, n^2))$ ancillae. Therefore, any such circuit of polynomial size $\mathcal{O}(n^c)$ can be parallelized to $\mathcal{O}(\log n)$ depth with $\mathcal{O}(kn^{c+1})$ ancillae, and any family of such circuits with fixed $k$ is in* $\mathbf{QNC}^1$.

*Proof.* Any such circuit can be written as the product of a circuit of controlled-not gates and a diagonal matrix that takes care of the phase shifts. The first part we can parallelize as in proposition 6, to $\mathcal{O}(\log n)$ depth and $\mathcal{O}(n^2)$ ancillae. As proposition 8 shows, diagonal matrices cannot be parallelized in general, so we have to look at the circuit more closely.

We can write the circuit we are trying to parallelize as a product $M = M_0 P_1 M_1 P_2 M_2 \cdots P_m M_m$ where the $M_i$ consist only of controlled-not gates and the $P_i$ are the diagonal operators. By passing the $P_i$ to the right end of the circuit, we can write

$$M = M_0 \cdots M_m \cdot D_1 \cdots D_m$$

where $D_i$ is the diagonal operator

$$D_i = (M_i \cdots M_m)^\dagger P_i (M_i \cdots M_m)$$

In other words, we simply calculate what state the controlled-not circuit was in when $P_i$ was applied, apply it, and uncalculate.

Each one of the $k$ qubits coupled by $P_i$ is the exclusive-or of some subset of the inputs, and can be calculated with a binary tree of $\mathcal{O}(n)$ ancillae as in proposition 6. Finally, by proposition 3 we can apply all the $D_i$ at once, by making $m$ copies of the system's entire state. Thus the total number of ancillae needed is $\mathcal{O}(kmn)$, or $\mathcal{O}(kn^{c+1})$ if $m = \mathcal{O}(n^c)$. □

## 8 The Hadamard gate, the Clifford group, and quantum codes

So far, all the circuits we have looked at are essentially classical; each row and each column has only one non-zero entry, so they are just reversible Boolean functions with phase shifts. Obviously, any interesting quantum algorithm will involve mixing between different Boolean states.

The simplest such operator is the *Hadamard gate* $R = \frac{1}{\sqrt{2}}\begin{pmatrix} 1 & 1 \\ 1 & -1 \end{pmatrix}$. By applying it all $n$ qubits of a state $|000\cdots 0\rangle$, we can prepare them in a superposition of all $2^n$ possible states. It is also the basic ingredient, along with phase shifts, of the standard circuit (shown below in figure 18) for the Quantum Fourier Transform.

We will call a controlled-$U$ gate a *controlled-Pauli gate* if $U$ is one of the Pauli matrices $\sigma_x = \begin{pmatrix} 0 & 1 \\ 1 & 0 \end{pmatrix}$, $-i\sigma_y = \begin{pmatrix} 0 & -1 \\ 1 & 0 \end{pmatrix}$, or $\sigma_z = \begin{pmatrix} 1 & 0 \\ 0 & -1 \end{pmatrix}$. Note that a controlled-$X$ is simply a controlled-not, a controlled-$Z$ is just the symmetric $\pi$-shift, and this real version of the controlled-$Y$ is their product.

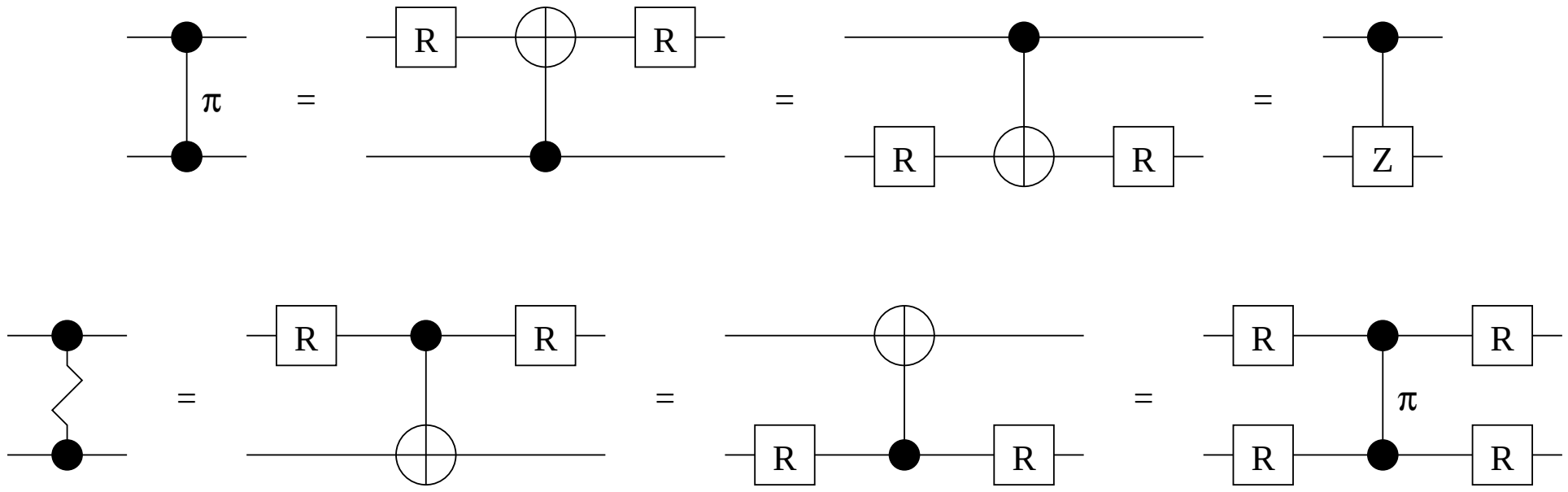

**Fig. 12.** Relations between the $\pi$-shift, the controlled-not, and the $w$ gate, which we notate with a wiggle.

The $\pi$-shift can be written in terms of a controlled-not by conjugating the target with $R$. Conjugating the input qubit instead gives us the $\pi$-shift in the Hadamard basis, which is a symmetric gate $\frac{1}{2}\begin{pmatrix} 1 & 1 & 1 & -1 \\ 1 & 1 & -1 & 1 \\ 1 & -1 & 1 & 1 \\ -1 & 1 & 1 & 1 \end{pmatrix}$. We call this the $w$-gate, and notate it as in figure 12.

Then we have the following:

**Proposition 10.** *Circuits of any size consisting of controlled-Pauli gates and the Hadamard gate $R$ can be parallelized to $\mathcal{O}(\log n)$ depth with $\mathcal{O}(n^2)$ ancillae. Thus any family of such circuits is in* $\mathbf{QNC}^1$.

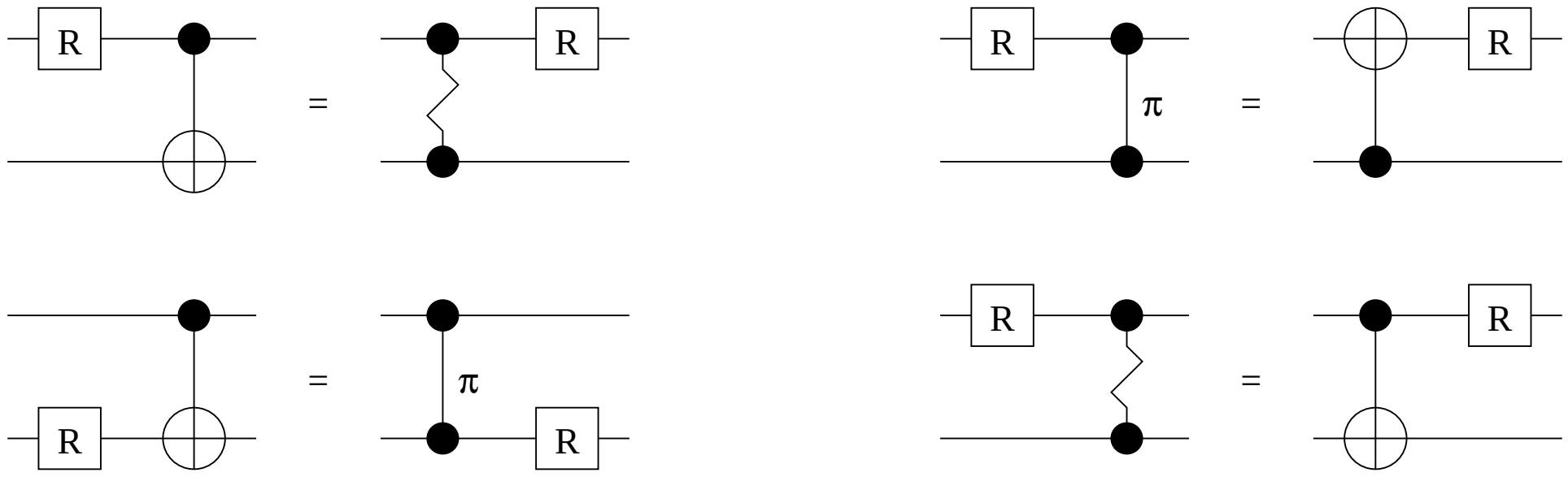

**Fig. 13.** Step 1: combing $R$'s to the right through controlled-nots, $\pi$-shifts, and $w$ gates.

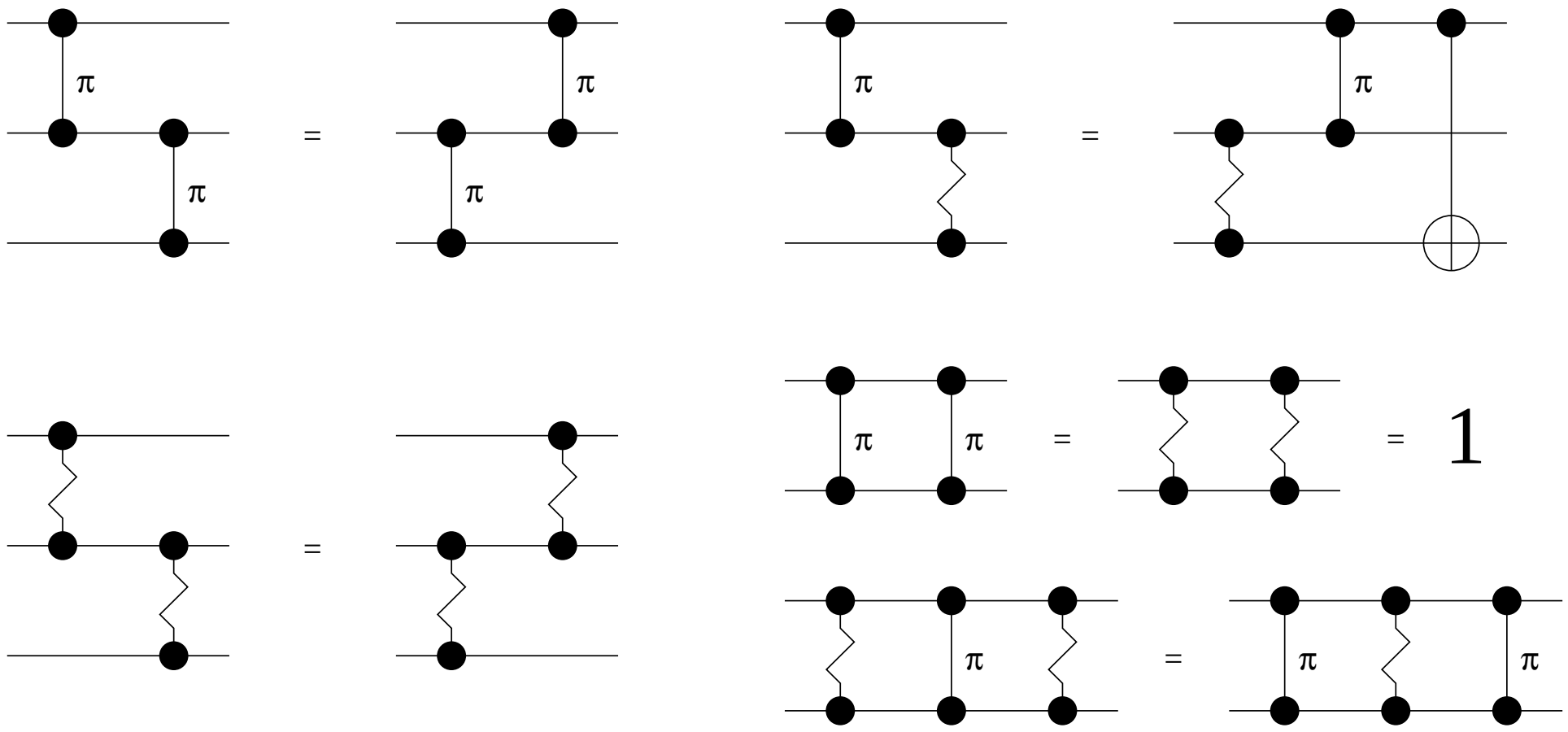

**Fig. 14.** Step 2: commuting $\pi$-shifts and $w$'s past each other, and combining them into 1, $z$, $w$, $zw$, $wz$, or $zwz$.

*Proof.* We will use the algebraic relations between these gates to arrange them into easily parallelizable groups. In step 1, we move Hadamard gates to the right through the other gates as shown in figure 13. This leaves a circuit of controlled-nots, $\pi$-shifts, and $w$-gates, followed by a single layer of $R$'s and identities.

In step 2, we arrange $\pi$-shifts and $w$-gates into three groups: a set of $\pi$-shifts, a set of $w$'s, and another set of $\pi$-shifts, with controlled-nots interspersed throughout. We can do this since when these gates are applied to different pairs of qubits, they commute (up to the creation of an additional controlled-not), and when applied to the same pair, generate a finite group. Specifically, if we call the $4 \times 4$ matrix of the $\pi$-shift $z$, then $z$ and $w$ obey the relations $z^2 = w^2 = 1$ and $wzw = zwz$, and generate the permutation group on three elements $S_3 = \{1, z, w, zw, wz, zwz\}$. Thus a group of $z$'s, a group of $w$'s, and a group of $z$'s are sufficient. These relations are shown in figure 14.

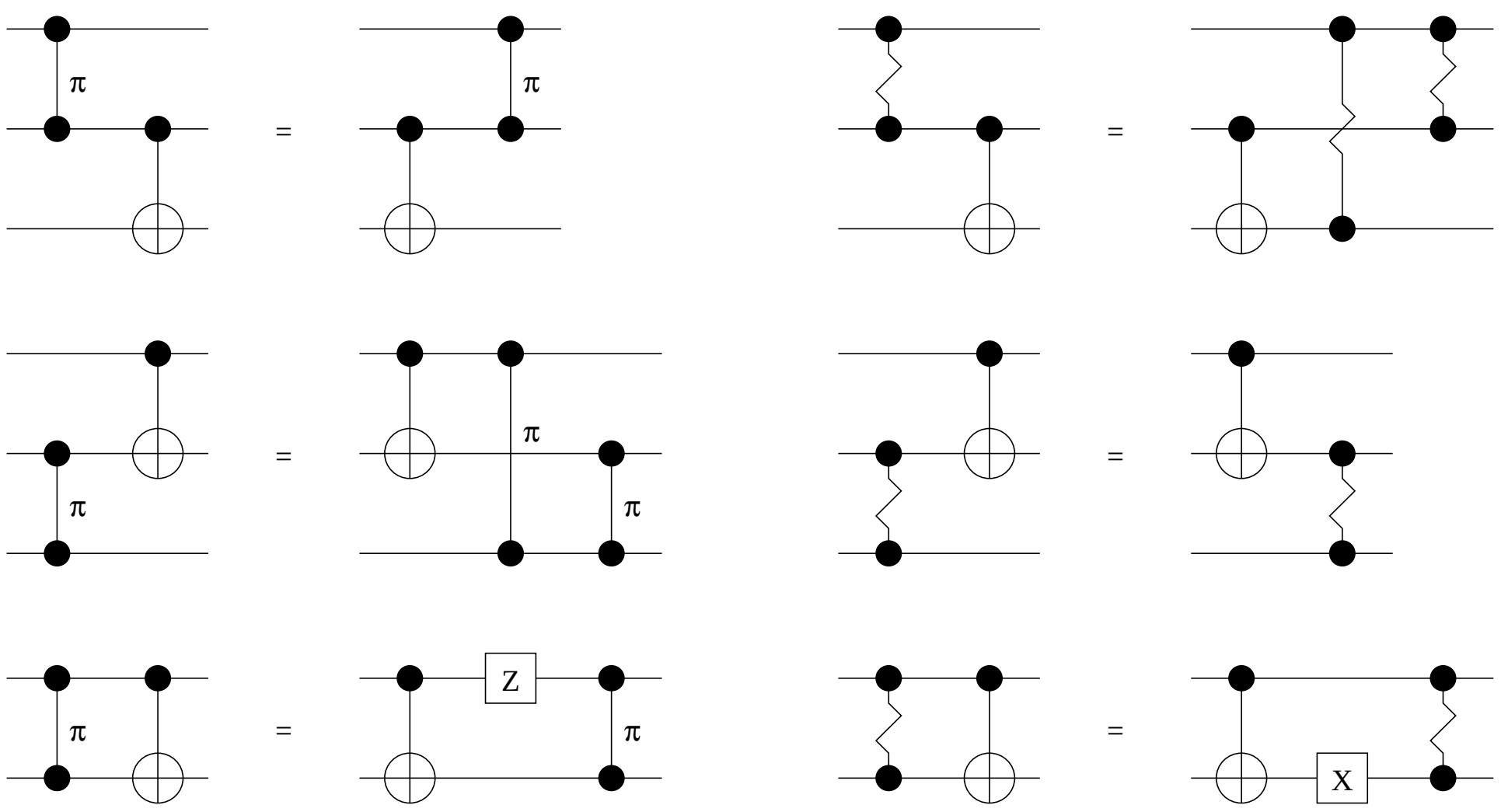

**Fig. 15.** Step 3: commuting $\pi$-shifts and $w$'s past controlled-nots.

In step 3, we pull the controlled-nots to the left through the $z$'s and $w$'s as shown in figure 15. This makes some additional symmetric gates, but always of the same type we pull through, so the grouping of $z$'s, $w$'s and $z$'s is not disturbed. (We also sometimes create single-qubit gates $X$ and $Z$, but these can be thought of as controlled-nots or $\pi$-shifts whose control qubit is always true.)

Finally, we note that since $w$ is simply $z$ in the Hadamard basis as shown in figure 12, we can write the group of $w$'s as a group of $z$'s conjugated with $R$ on every qubit. We are left with a circuit of controlled-not gates, followed by three groups of $\pi$-shifts separated by two layers of $R$'s, and a single layer of possible $R$'s as shown schematically in figure 16.

Propositions 5 and 6 show how to parallelize circuits of $\pi$-shifts and of controlled-nots to $\mathcal{O}(\log n)$ depth with $\mathcal{O}(n^2)$ ancillae, and the theorem is proved. $\square$

With a little extra work we should also be able to include the one-qubit $\pi/2$ shift $P = \begin{pmatrix} 1 & 0 \\ 0 & i \end{pmatrix}$. This would give us the *Clifford group*, which is the normalizer of the group of tensor products of Pauli matrices. In fact, some of the relations we have used here are equivalent to those used by Gottesman to derive the Heisenberg representation of circuits in the Clifford group [7].

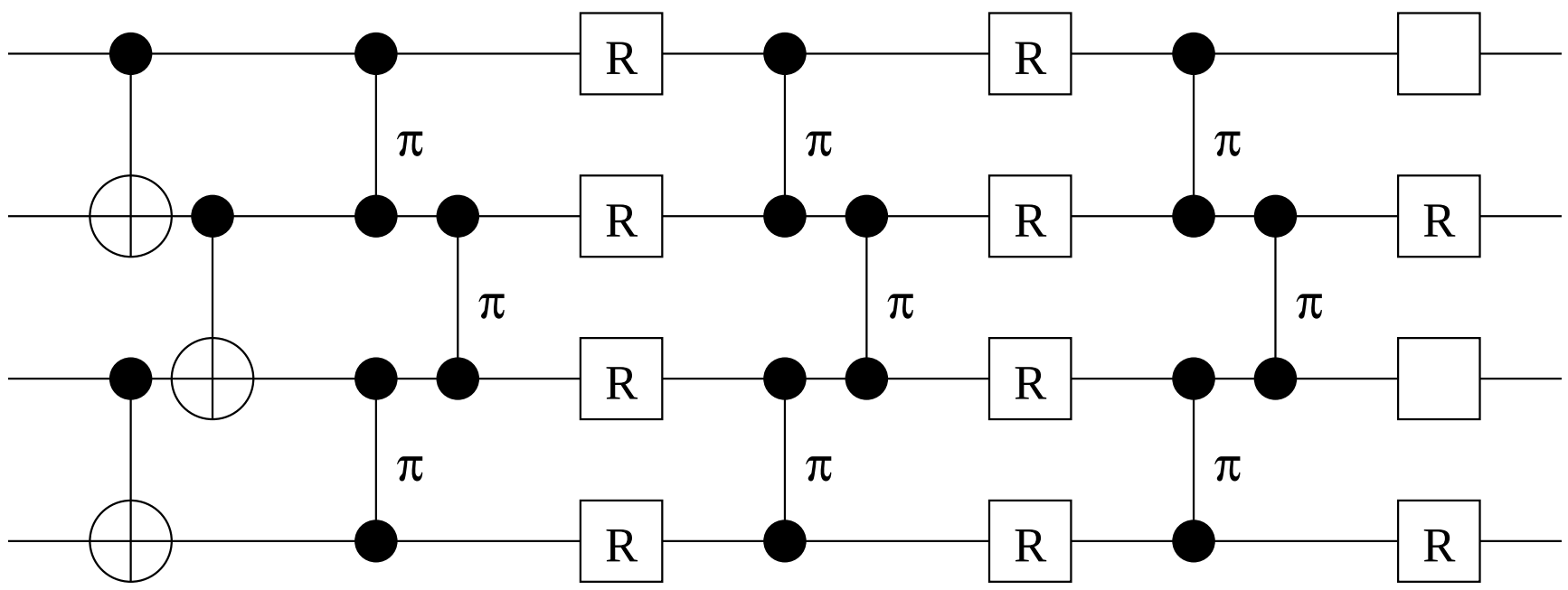

**Fig. 16.** The kind of circuit we are left with after steps 1, 2, and 3, and after writing the $w$'s as $\pi$-shifts conjugated by $R$.

There may be other interesting finite subgroups of $O(2^n)$ that we can parallelize. However, if we add the two-qubit controlled-$P$ gate (also known as the 'square-root-of-not') we get universal computation, i.e. we can generate a dense set of quantum operators. Algebraically, this shows up as the fact that the controlled-$Z$ gate is the only two-qubit phase shift whose conjugate by a controlled-not can be expressed with two- and one-qubit gates, just as $P$ and $Z$ are the only one-qubit phase shifts whose conjugate by a controlled-not can be expressed with themselves and controlled-$Z$ gates. Other phase shifts generate three- and more-qubit interactions when they are commuted through controlled-nots.

In any case, this gives us the following corollary.

**Corollary 1.** *Additive (or 'stabilizer') quantum error-correcting codes are in* $\mathbf{QNC}^1$*, in the sense that encoding and decoding families of such codes with* $n$*-qubit code words can be done in* $\mathcal{O}(\log n)$ *depth and* $\mathcal{O}(n^2)$ *ancillae.*

*Proof.* Since the Pauli matrices $\sigma_x$ and $\sigma_z$ generate bit errors and phase errors respectively, circuits for quantum codes such as those in [15,9,2,5] are composed of controlled-Pauli and Hadamard gates. By a result of Rains [13], additive quantum codes are always equivalent to real ones, so the real version of the controlled-$Y$ gate is sufficient. □

In fact, Cleve and Gottesman [3] and Steane [16] have shown that circuits for additive quantum codes can be constructed out of controlled-Pauli gates, where Hadamard gates appear only in one or two layers. Thus proposition 9 is already enough to parallelize these circuits.

## 9 QNC ≠ QP? The staircase circuit

A simple, perhaps minimal, example of a quantum circuit that seems hard to parallelize is the "staircase" circuit shown in figure 17. This kind of structure appears in the standard circuit for the quantum Fourier transform, which has $\mathcal{O}(n^2)$ gates [4,14]. Careful inspection shows that the QFT can in fact be parallelized to $\mathcal{O}(n)$ depth as shown in figure 18 (an upside-down version of which is given in [8]), but it seems difficult to do any better. Clearly, any fast parallel circuit for the QFT would be relevant to prime factoring and other problems the QFT is used for.

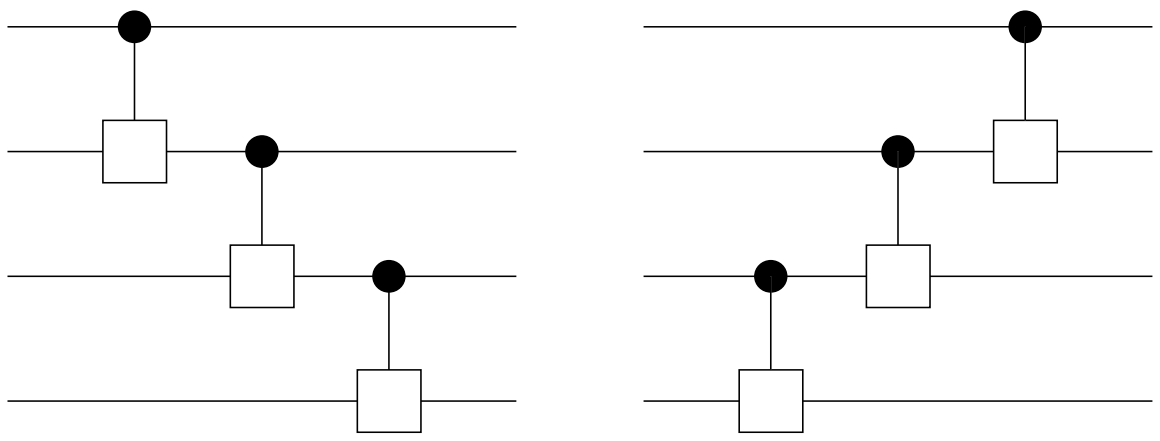

**Fig. 17.** These "staircase" circuits seems hard to parallelize unless the operators are purely diagonal or off-diagonal.

If we define **QP** as the family of quantum operators that can be expressed with circuits of polynomial depth (again, leaving measurement issues aside for now), we can make the following conjecture:

*Conjecture 1.* Staircase circuits composed of controlled-$U$ gates other than diagonal or off-diagonal gates (i.e. other than the special cases handled in propositions 5 and 6) cannot be parallelized to less than linear depth. Therefore, $\mathbf{QNC} < \mathbf{QP}$.

## 10 Conclusion

We conclude with some questions for further work.

Does parallelizing the encoding and decoding of error-correcting codes help reduce the error threshold for reliable quantum computation, at least in regimes where storage errors are more significant than gate errors?

Parsing classical context-free languages is in **NC**, and quantum context-free languages have been defined in [10]. Is quantum parsing, i.e. producing derivation trees with the appropriate amplitudes, in **QNC**?

Finally, can the reader show that the staircase circuit cannot be parallelized, thus showing that $\mathbf{QNC} < \mathbf{QP}$? This would be quite significant, since corresponding classical question $\mathbf{NC} < \mathbf{P}$ is still open, and believed to be very hard.

**Acknowledgments.** M.N. would like to thank the Santa Fe Institute for their hospitality, and Spootie the Cat for her affections. C.M. would like to thank the organizers of the First International Conference on Unconventional Models of Computation in Auckland, New Zealand, as well as Seth Lloyd, Tom Knight, David DiVincenzo, Artur Ekert, Jonathan Machta, Mike Nielsen, and Daniel Gottesman for helpful conversations. He would also like to thank Molly

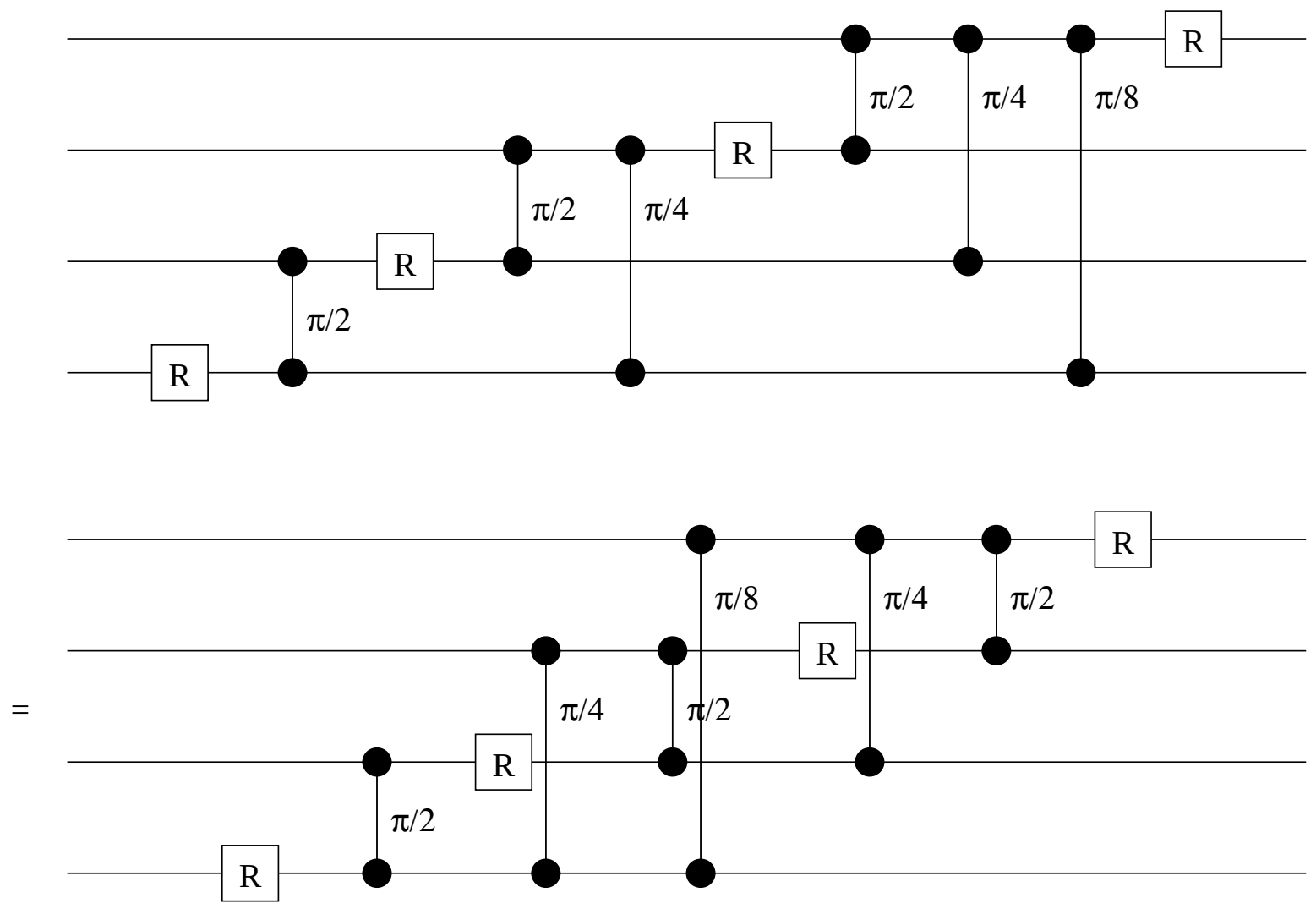

**Fig. 18.** The standard circuit for the quantum Fourier transform on $n$ qubits can be carried out in $2n-1$ layers. Can it be parallelized to less than linear depth?

Rose for inspiration and companionship. This work was supported in part by NSF grant ASC-9503162.